\documentclass[12pt,english]{article}
\usepackage[scaled=0.920000017]{helvet}

\usepackage[T1]{fontenc}
\usepackage[latin9]{inputenc}
\usepackage{geometry}
\usepackage{verbatim}
\usepackage{float}
\usepackage{amsmath}
\usepackage{graphicx}
\usepackage{setspace}
\usepackage{esint}
\usepackage{scrextend}

\usepackage{hyperref}

\hypersetup{pdfauthor={Brito \& Gerstner},pdftitle={Nonlinear Hebbian learning}}

\usepackage[space]{grffile}

\makeatletter
\numberwithin{figure}{section}
 
\usepackage{chngcntr}
\counterwithout{figure}{section}

\usepackage[super,numbers,sort&compress]{natbib}

\makeatother

\usepackage{babel}
\begin{document}

{\centering
\textbf{\Large { Nonlinear Hebbian learning as a unifying principle \\ in receptive field formation}}

\vspace{0.5cm}

Carlos S. N. Brito*, Wulfram Gerstner

\vspace{0.3cm}

{\singlespacing
School of Computer and Communication Sciences and School of Life Sciences, \\ 
Ecole Polytechnique Federale de Lausanne (EPFL), Switzerland \\
*Corresponding author: carlos.stein@epfl.ch 
}

\vspace{1.0cm}

\textbf{\large {Abstract}}

\vspace{0.6cm}

\par
}

\begin{addmargin}[4em]{4em}
\noindent The development of sensory receptive fields has
been modeled in the past by a variety of models including normative
models such as sparse coding or independent component analysis and
bottom-up models such as spike-timing dependent plasticity or the
Bienenstock-Cooper-Munro model of synaptic plasticity. Here we show
that the above variety of approaches can all be unified into a single
common principle, namely Nonlinear Hebbian Learning. When Nonlinear
Hebbian Learning is applied to natural images, receptive field shapes
were strongly constrained by the input statistics and preprocessing,
but exhibited only modest variation across different choices of nonlinearities
in neuron models or synaptic plasticity rules. Neither overcompleteness
nor sparse network activity are necessary for the development of localized
receptive fields. The analysis of alternative sensory modalities such
as auditory models or V2 development lead to the same conclusions.
In all examples, receptive fields can be predicted a priori by reformulating
an abstract model as nonlinear Hebbian learning. Thus nonlinear Hebbian
learning and natural statistics can account for many aspects of receptive
field formation across models and sensory modalities. 
\end{addmargin}



\pagebreak{}

\vspace{1cm}

\section*{Introduction}

Neurons in sensory areas of the cortex are optimally driven by stimuli
with characteristic features that define the 'receptive field' of
the cell. While receptive fields of simple cells in primary visual
cortex (V1) are localized in visual space and sensitive to the orientation
of light contrast \citep{hubel_receptive_1959}, those of auditory neurons
are sensitive to specific time-frequency patterns in sounds \citep{Miller_Escabi_Read_Schreiner_2002}.
The concept of a receptive field is also useful when studying higher-order
sensory areas, for instance when analyzing the degree of selectivity
and invariance of neurons to stimulus properties \citep{DiCarlo_Zoccolan_Rust_2012,Freeman_Simoncelli_2011}.

The characteristic receptive fields of simple cells in V1 have been
related to statistical properties of natural images \citep{Field_1994}.
These findings inspired various models, based on principles as diverse
as sparse sensory representations \citep{olshausen_emergence_1996},
optimal information transmission \citep{bell_independent_1997}, or
synaptic plasticity \citep{law_formation_1994}. Several studies highlighted
possible connections between biological and normative justifications
of sensory receptive fields \citep{Rehn_Sommer_2007,clopath_connectivity_2010,Savin_Joshi_Triesch_2010,zylberberg_sparse_2011},
not only in V1, but also in other sensory areas \citep{olshausen_sparse_2004},
such as auditory \citep{Smith_Lewicki_2006,Saxe_Bhand_Mudur_Suresh_Ng_2011}
and secondary visual cortex (V2)  \citep{Lee_Ekanadham_Ng_2007}. 

Since disparate models appear to achieve similar results, the question
arises whether there exists a general underlying concept in unsupervised
learning models \citep{Saxe_Bhand_Mudur_Suresh_Ng_2011,yamins_performance-optimized_2014}.
Here we show that the principle of nonlinear Hebbian learning is sufficient
for receptive field development under rather general conditions. The
nonlinearity is defined by the neuron's f-I curve combined with the
nonlinearity of the plasticity function. The outcome of such nonlinear
learning is equivalent to projection pursuit \citep{Friedman_1987,Oja_Ogawa_Wangviwattana_1991,Fyfe_Baddeley_1995},
which focuses on features with non-trivial statistical structure,
and therefore links receptive field development to optimality principles.

Here we unify and broaden the above concepts and show that plastic
neural networks, sparse coding models and independent component analysis
can all be reformulated as nonlinear Hebbian learning. For natural
images as sensory input, we find that a broad class of nonlinear Hebbian
rules lead to orientation selective receptive fields, and explain
how seemingly disparate approaches may lead to similar receptive fields.
The theory predicts the diversity of receptive field shapes obtained
in simulations for several different families of nonlinearities. The
robustness to model assumptions also applies to alternative sensory
modalities, implying that the statistical properties of the input
strongly constrain the type of receptive fields that can be learned.
Since the conclusions are robust to specific properties of neurons
and plasticity mechanisms, our results support the idea that synaptic
plasticity can be interpreted as nonlinear Hebbian learning, implementing
a statistical optimization of the neuron's receptive field properties.

\section*{Results }

\subsection*{The effective Hebbian nonlinearity}

\begin{figure}[!htb]
\begin{centering}
\includegraphics{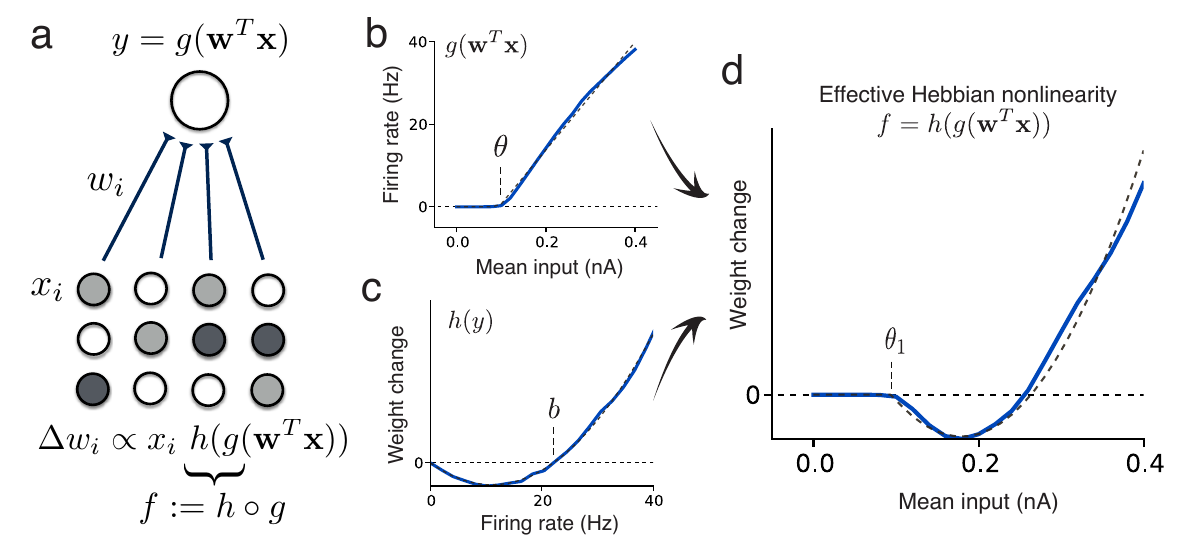}
\par\end{centering}

\caption[The effective Hebbian nonlinearity of plastic cortical networks.]{\textbf{The effective Hebbian nonlinearity of plastic cortical networks.} (\textbf{a})
Receptive field development between an input layer of neurons with
activities $x_{i}$, connected by synaptic projections $w_{i}$ to
a neuron with firing rate $y$, given by an f-I curve $y=g(\mathbf{w}^{T}\mathbf{x}))$.
Synaptic connections change according to a Hebbian rule $\Delta w_{i}\propto x_{i}\ h(y)$.
(\textbf{b}) f-I curve (blue) of a GIF model \citep{pozzorini_temporal_2013}
of a pyramidal neurons in response to step currents of $\mbox{500}$
ms duration (dashed line: piece-wise linear fit, with slope $a=143$
Hz/nA and threshold $\theta=0.08$ nA). (\textbf{c}) Plasticity function
of the triplet STDP model \citep{pfister_triplets_2006} (blue), fitted
to visual cortex plasticity data \citep{Sjostrom_Turrigiano_Nelson_2001,pfister_triplets_2006},
showing the weight change $\Delta w_{i}$ as a function of the post-synaptic
rate $y$, under a constant pre-synaptic stimulation $x_{i}$ (dashed
line: fit by quadratic function, with LTD factor $b=22.1$ Hz). (\textbf{d})
The combination of the f-I curve and plasticity function generates
the effective Hebbian nonlinearity (dashed line: quadratic nonlinearity
with LTD threshold $\theta_{1}=0.08$ nA, LTP threshold $\theta_{2}=0.23$
nA). }
\label{fig:intro}
\end{figure}

In classic rate models of sensory development \citep{Miller_Keller_Stryker_1989,law_formation_1994,olshausen_emergence_1996},
a first layer of neurons, representing the sensory input $\mathbf{x}$,
is connected to a downstream neuron with activity $y$, through synaptic
connections with weights $\mathbf{w}$ (Fig. \ref{fig:intro}a).
The response to a specific input is $y=g(\mathbf{w}^{T}\mathbf{x})$,
where $g$ is the frequency-current (f-I) curve. In most models of
Hebbian plasticity \citep{Bienenstock_Cooper_Munro_1982,Gerstner_Kistler_Naud_Paninski_2014},
synaptic changes $\Delta\mathbf{w}$ of the connection weights depend
on pre- and post-synaptic activity, with a linear dependence on the
pre-synaptic and a nonlinear dependence on the post-synaptic activity,
$\Delta\mathbf{w}\propto\mathbf{x}\ h(y)$, in accordance with models
of pairing experiments \citep{pfister_triplets_2006,clopath_connectivity_2010}.
The learning dynamics arise from a combination of the neuronal f-I
curve $y=g(\mathbf{w^{T}x})$ and the Hebbian plasticity function
$\Delta\mathbf{w}\propto\mathbf{x}\ h(y)$:

\begin{align}
\Delta\mathbf{w} & \propto\mathbf{x}\ h(g(\mathbf{w^{T}x}))=\mathbf{x}\ f(\mathbf{w^{T}x})\label{fgh}
\end{align}
where we define the \emph{effective Hebbian nonlinearity} $f:=h\circ g$
as the composition of the nonlinearity in the plasticity rule and
the neuron's f-I curve. In an experimental setting, the pre-synaptic
activity $x$ is determined by the set of sensory stimuli (influenced
by, e.g., the rearing conditions during sensory development \citep{Wiesel_Hubel_others_1963}).
Therefore, the evolution of synaptic strength, Eq. \ref{fgh},
is determined by the effective nonlinearity $f$ and the statistics
of the input $\mathbf{x}$. 

Many existing models can be formulated in the framework of Eq.
\ref{fgh}. For instance, in a classic study of simple-cell formation
\citep{law_formation_1994}, the Bienenstock-Cooper-Munro (BCM) model
\citep{Bienenstock_Cooper_Munro_1982} has a quadratic plasticity nonlinearity,
$h(y)=y(y-\theta)$, with a variable plasticity threshold $\theta$,
and a sigmoidal f-I curve, $\sigma(\mathbf{w}^{T}\mathbf{x})$, which
combine into nonlinear Hebbian learning dynamics, $\Delta\mathbf{w}\propto\mathbf{x}\ h(\sigma(\mathbf{w}^{T}\mathbf{x}))$. 

More realistic cortical networks have dynamical properties which are
not accounted for by rate models. By analyzing state-of-the-art models
of cortical neurons and synaptic plasticity, we inspected whether
plastic spiking networks can be reduced to nonlinear Hebbian learning.
We considered a generalized leaky integrate-and-fire model (GIF),
which includes adaptation, stochastic firing and predicts experimental
spikes with high accuracy \citep{pozzorini_temporal_2013}, and we
approximate its f-I curve by a linear rectifier, $g(u)=a(u-\theta)_{+}$,
with slope $a$ and threshold $\theta$ (Fig. \ref{fig:intro}b). 

As a phenomenological model of synaptic plasticity grounded on experimental
data \citep{Sjostrom_Turrigiano_Nelson_2001}, we implemented triplet
spike-timing dependent plasticity (STDP) \citep{pfister_triplets_2006}.
In this STDP model, the dependence of long-term potentiation (LTP)
upon two post-synaptic spikes induces in the corresponding rate model
a quadratic dependence on the post-synaptic rate, while long-term
depression (LTD) is linear. The resulting rate plasticity \citep{pfister_triplets_2006}
is $h(y)=y^{2}-by$, with an LTD factor $b$ (post-synaptic activity
threshold separating LTD from LTP, Fig. \ref{fig:intro}c),
similar to the classic BCM model \citep{Bienenstock_Cooper_Munro_1982,law_formation_1994}.

Composing the f-I curve of the GIF with the $h(y)$ for the triplet
plasticity model, we have an approximation of the effective learning
nonlinearity $f=h\circ g$ in cortical spiking neurons (Fig
\ref{fig:intro}d), that can be described as a quadratic rectifier,
with LTD threshold given by $\theta_{1}=\theta$ and LTP threshold
given by $\theta_{2}=\theta+b/a$. Interestingly, the f-I slope $a$
and LTD factor $b$ are redundant parameters of the learning dynamics:
only their ratio counts in nonlinear Hebbian plasticity. Metaplasticity
can control the LTD factor \citep{pfister_triplets_2006,turrigiano_too_2011},
thus regulating the LTP threshold.

If one considers a linear STDP model \citep{Song_Miller_Abbott_2000,Gerstner_Kempter_Van_Hemmen_1996}
instead of the triplet STDP \citep{pfister_triplets_2006}, the plasticity
curve is linear \citep{Gerstner_Kistler_Naud_Paninski_2014}, as in
standard Hebbian learning, and the effective nonlinearity is shaped
by the properties of the f-I curve (Fig.  \ref{fig:winners}a). 

\subsection*{Sparse coding as nonlinear Hebbian learning}

Beyond phenomenological modeling, normative principles that explain
receptive fields development have been one of the goals of theoretical
neuroscience \citep{dayan_theoretical_2001}. Sparse coding \citep{olshausen_emergence_1996}
starts from the assumptions that V1 aims at maximizing the sparseness
of the activity in the sensory representation, and became a well-known
normative model to develop orientation selective receptive fields
\citep{Rehn_Sommer_2007,zylberberg_sparse_2011,olshausen_sparse_2004}.
We demonstrate that the algorithm implemented in the sparse coding
model is in fact a particular example of nonlinear Hebbian learning.

The sparse coding model aims at minimizing an input reconstruction
error $E=\frac{1}{2}||\mathbf{x}-\mathbf{W}\mathbf{y}||^{2}+\lambda S(\mathbf{y})$,
under a sparsity constraint $S$ with relative importance $\lambda>0$.
For $K$ hidden neurons $y_{j},$ such a model implicitly assumes
that the vector $\mathbf{w_{j}}$ of feed-forward weights onto neuron
$j$ are mirrored by hypothetical \textquotedbl{}reconstruction weights\textquotedbl{},
$\mathbf{W}=[\mathbf{w}_{1}\dots\mathbf{w}_{K}]$. The resulting encoding
algorithm can be recast as a neural model \citep{rozell_sparse_2008},
if neurons are embedded in a feedforward model with lateral inhibition,
$\mathbf{y}=g(\mathbf{w}^{T}\mathbf{x}-\mathbf{v}^{T}\mathbf{y})$,
where $v$ are inhibitory recurrent synaptic connections (see Methods).
In the case of a single output neuron, its firing rate is simply $y=g(\mathbf{w}^{T}\mathbf{x})$.
The nonlinearity $g$ of the f-I curve is threshold-like, and determined
by the choice of the sparsity constraint \citep{rozell_sparse_2008},
such as the Cauchy, $L_{0}$ , or $L_{1}$ constraints (Fig
\ref{fig:winners}a, see Methods).

If weights are updated through gradient descent so as to minimize
$E$, the resulting plasticity rule is Oja's learning rule \citep{Oja_1982},
$\Delta\mathbf{w}\propto\mathbf{x}\ y-\mathbf{w}\ y^{2}$. The second
term $-\mathbf{w}\ y^{2}$ has a multiplicative effect on the strength
of synapses projecting onto the same neuron (weight rescaling), but
does not affect the receptive field shape, whereas the first term
$\mathbf{x}\ y$ drives feature selectivity and receptive field formation.
Together, these derivations imply that the one-unit sparse coding
algorithm can be implemented by an effective nonlinear Hebbian rule
combined with weight normalization. Although the plasticity mechanism
is linear, $\Delta\mathbf{w}\propto\mathbf{x}\ y$, a nonlinearity
arises from the f-I curve, $y=g(\mathbf{w}^{T}\mathbf{x})$, so that
the effective plasticity is
\begin{align}
\Delta\mathbf{w} & \propto\mathbf{x}\ g(\mathbf{w}^{T}\mathbf{x})\label{SCfgh}
\end{align}

This analysis reveals an equivalence between sparse coding models
and neural networks with linear plasticity mechanisms, where the sparsity
constraint is determined by the f-I curve $g$.

Similarly, algorithms performing independent component analysis (ICA),
a model class closely related to sparse coding, also perform effective
nonlinear Hebbian learning, albeit inversely, with linear neurons
and a nonlinear plasticity rule \citep{hyvarinen_independent_2000}.
For variants of ICA based on information maximization \citep{bell_independent_1997}
or kurtosis \citep{hyvarinen_independent_2000} different nonlinearities
arise (Fig. \ref{fig:winners}a), but Eq. \ref{SCfgh}
applies equally well. Hence, various instantiations of sparse coding
and ICA models not only relate to each other in their normative assumptions
\citep{olshausen_sparse_1997}, but when implemented as iterative gradient
update rules, they all employ nonlinear Hebbian learning.

\subsection*{Simple cell development for a large class of nonlinearities}

Since the models described above can be implemented by similar plasticity
rules, we hypothesized nonlinear Hebbian learning to be a general
principle that explains the development of receptive field selectivity.
Nonlinear Hebbian learning with an effective nonlinearity $f$ is
linked to an optimization principle with a function $F=\int f$ \citep{Oja_Ogawa_Wangviwattana_1991,Fyfe_Baddeley_1995}.
For an input ensemble $\mathbf{x}$, optimality is achieved by weights
$\tilde{\mathbf{w}}$ that maximize $\langle F(\tilde{\mathbf{w}}{}^{T}\mathbf{x})\rangle$,
where angular brackets denote the average over the input statistics.
Nonlinear Hebbian learning is a stochastic gradient ascent implementation
of this optimization process, called projection pursuit \citep{Friedman_1987,Oja_Ogawa_Wangviwattana_1991,Fyfe_Baddeley_1995}:

\begin{equation}
\tilde{\mathbf{w}}=max_{\mathbf{w}}\langle F(\mathbf{w}^{T}\mathbf{x})\rangle\implies\Delta\mathbf{w}\propto\mathbf{x}\ f(\mathbf{w}^{T}\mathbf{x})
\end{equation}

\begin{figure}[!htbp]
\begin{centering}
\includegraphics[height=0.6\textheight]{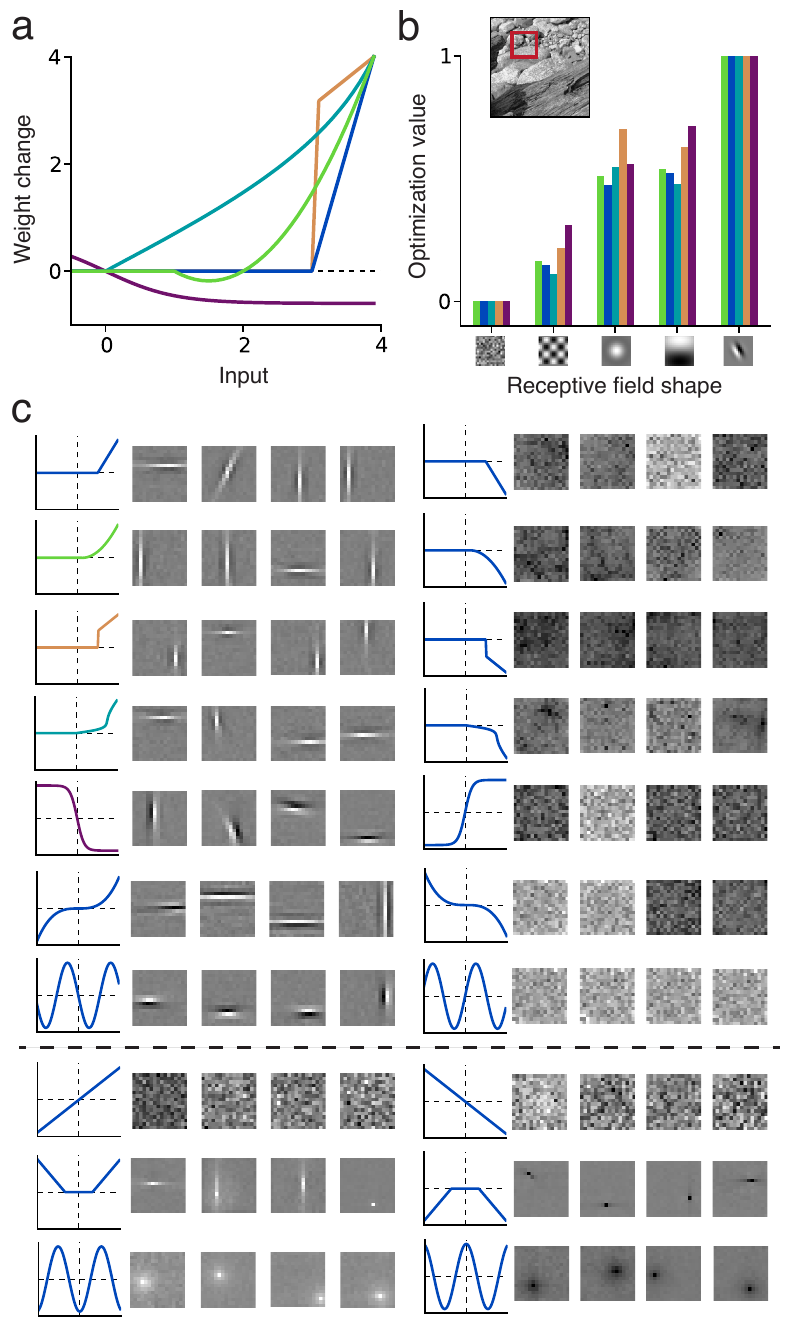}
\par\end{centering}

\caption[Simple cell development from natural images regardless of specific
effective Hebbian nonlinearity.]{\textbf{Simple cell development from natural images regardless of specific
effective Hebbian nonlinearity.} (\textbf{a}) Effective nonlinearity
of five common models (arbitrary units): quadratic rectifier (green,
as in cortical and BCM models, $\theta_{1}=1.$, $\theta_{2}=2.$),
linear rectifier (dark blue, as in $L_{1}$ sparse coding or networks
with linear STDP, $\theta=3.$), Cauchy sparse coding nonlinearity
(light blue, $\lambda=3.$), $L_{0}$ sparse coding nonlinearity (orange,
$\lambda=3.$), and negative sigmoid (purple, as in ICA models). (\textbf{b})
Relative optimization value $\langle F(\mathbf{w}^{T}\mathbf{x})\rangle$
for each of the five models in \textbf{a}, for different preselected
features $\mathbf{w}$, averaged over natural image patches \textbf{$\mathbf{x}$}.
Candidate features are represented as two-dimensional receptive fields.
For all models, the optimum is achieved at the localized oriented
receptive field. Inset: Example of natural image and image patch (red
square) used as sensory input. (\textbf{c}) Receptive fields learned
in four trials for ten effective Hebbian functions $f$ (from top:
the five functions considered above, $u^{3}$, $-sin(u)$, $u$, $(|u|-2)_{+}$,
$-cos(u)$)\textbf{ }(\textbf{left} \textbf{column}), and their opposites
$-f$ (\textbf{right column}). The first seven functions (above the
dashed line) lead to localized oriented filters, while a sign-flip
leads to random patterns. Linear or symmetric functions are exceptions
and do not develop oriented filters (\textbf{bottom} \textbf{rows}).}

\label{fig:winners}
\end{figure}

Motivated by results from ICA theory \citep{hyvarinen_independent_1998}
and statistical properties of whitened natural images \citep{Field_1994},
we selected diverse Hebbian nonlinearities $f$ (Fig.  \ref{fig:winners}a)
and calculated the corresponding optimization value $\langle F(\mathbf{w}^{T}\mathbf{x})\rangle$
for different features of interest that we consider as candidate RF
shapes, with a whitened ensemble of patches extracted from natural
images as input (see Methods). These include a random connectivity
pattern, a non-local oriented edge (as in principal components of
natural images) and localized oriented edges (as in cat and monkey
simple cells in the visual cortex), shown in Fig. \ref{fig:winners}b.
The relative value of $\langle F(\mathbf{w}^{T}\mathbf{x})\rangle$
between one feature and another was remarkably consistent across various
choices of the nonlinearity $f$, with localized orientation-selective
receptive fields as maxima (Fig.  \ref{fig:winners}b). Furthermore,
we also searched for the maxima through gradient ascent, so as to
confirm that the maxima are orientation selective (Fig.  \ref{fig:winners}c,
left). Our results indicate that receptive field development of simple
cells is mainly governed by the statistical properties of natural
images, while robust to specific model assumptions. 

\begin{figure}[!htbp]
\centering{}\includegraphics{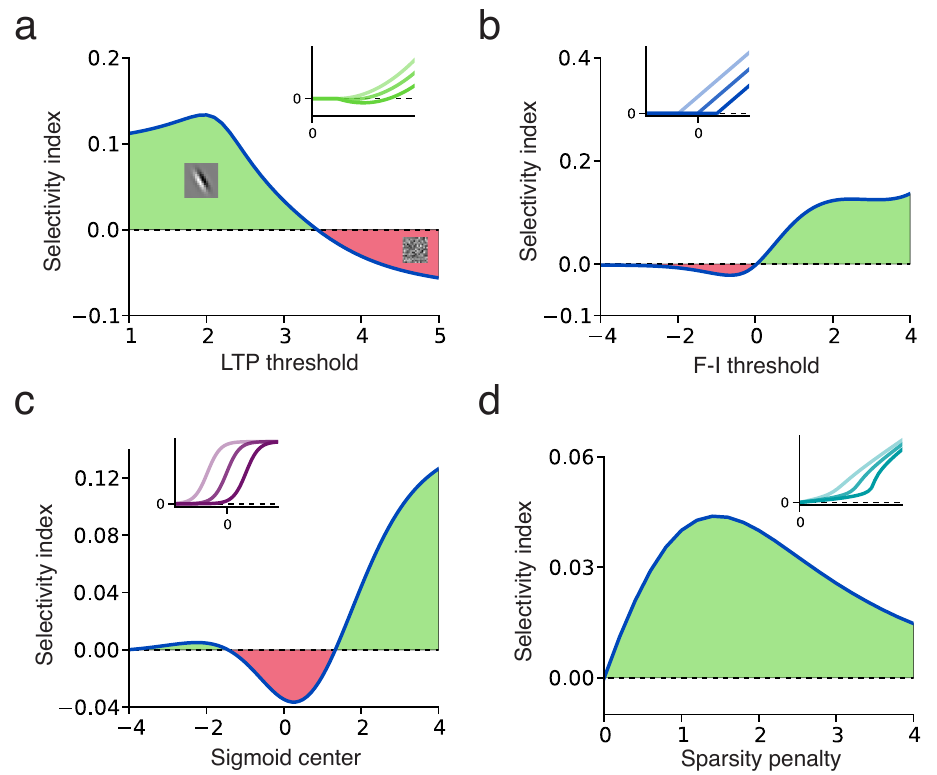}
\caption[Selectivity index for different effective nonlinearities.]
{\textbf{Selectivity index for different effective nonlinearities.} (\textbf{a})
Quadratic rectifier (small graphic, three examples with different
LTP thresholds) with LTD threshold at $\theta_{1}=1$: LTP threshold
must be below $3.5$ to secure positive selectivity index (green region,
main Fig) and learn localized oriented receptive fields (inset). A
negative selectivity index (red region) leads to a random connectivity
pattern (inset) (\textbf{b}) Linear rectifier: activation threshold
must be above zero. (\textbf{c}) Sigmoid: center must be below $a=-1.2$
or, for a stronger effect, above $a=+1.2$. The opposite conditions
apply to the negative sigmoid. (\textbf{d}) Cauchy sparse coding nonlinearity:
positive but weak feature selectivity for any sparseness penalty $\lambda>0$.
Insets show the nonlinearities for different choices of parameters. }
\label{fig:selectivity}
\end{figure}

The relevant property of natural image statistics is that the distribution
of a feature derived from typical localized oriented patterns has
high kurtosis \citep{Field_1994,olshausen_emergence_1996,ruderman_statistics_1994}.
Thus to establish a quantitative measure whether a nonlinearity is
suitable for feature learning, we define a \emph{selectivity index}
(\emph{SI}), which measures the relative value of $\langle F(.)\rangle$
between a variable $l$ with a Laplacian distribution and a variable
$g$ with Gaussian distribution \citep{hyvarinen_independent_1998}:
$SI=(\langle F(l)\rangle-\langle F(g)\rangle)/\sigma_{F}$ (see Methods).
The Laplacian variable has higher kurtosis than the Gaussian variable,
serving as a prototype of a kurtotic distribution. Since values obtained
by filtering natural images with localized oriented patterns have
a distribution with longer tails than other patterns \citep{Field_1994},
as does the Laplacian variable compared to the Gaussian, positive
values $SI>0$ indicate good candidate functions for learning simple
cell-like receptive fields from natural images. We find that each
model has an appropriate parameter range where $SI>0$ (Fig.
 \ref{fig:selectivity}). For example the quadratic rectifier nonlinearity
needs an LTP threshold $\theta_{2}$ below some critical level, so
as to be useful for feature learning (Fig. \ref{fig:selectivity}a).

A sigmoidal function with threshold at zero has \emph{negative SI},
but a \emph{negative} sigmoid, as used in ICA studies \citep{bell_independent_1997},
has $SI>0$. More generally, whenever an effective nonlinearity $f$
is not suited for feature learning, its opposite $-f$ should be,
since its $SI$ will have the opposite sign (Fig. \ref{fig:winners}c).
This implies that, in general, half of the function space could be
suitable for feature learning \citep{hyvarinen_independent_1998},
i.e. it finds weights $w$ such that the distribution of the feature
$\mathbf{w}^{T}\mathbf{x}$ has a long tail, indicating high kurtosis
(\textquotedbl{}kurtotic feature\textquotedbl{}). The other half of
the function space learns the least kurtotic features (e.g. random
connectivity patterns for natural images, Fig. \ref{fig:winners}b,c). 

This universality strongly constrains the possible shape of receptive
fields that may arise during development for a given input dataset.
For whitened natural images, a learnable receptive field is in general
either a localized edge detector or a non-localized random connectivity
pattern. 

An important special case is an effective linear curve, $f(u)=u$,
which arises when both f-I and plasticity curves are linear \citep{Miller_Keller_Stryker_1989}.
Because the linear model maximizes variance $\langle(\mathbf{w}^{T}\mathbf{x})^{2}\rangle$,
it can perform principal component analysis \citep{Oja_1982}, but
does not have any feature selectivity on whitened input datasets,
where variance is constant (Fig. \ref{fig:winners}c). 

Symmetric effective nonlinearities, $f(u)=f(-u)$, are also exceptions,
since their corresponding optimization functions are asymmetric, $F(u)=-F(-u)$,
so that for datasets with symmetric statistical distributions, $P(\mathbf{x})=P(-\mathbf{x})$,
the optimization value will be zero, $\langle F_{asym.}(\mathbf{w}^{T}\mathbf{x}_{sym.})\rangle=0$.
As natural images are not completely symmetric, localized receptive
fields do develop, though without orientation selectivity, as illustrated
by a cosine function and a symmetric piece-wise linear function as
effective nonlinearities (Fig. \ref{fig:winners}c, bottom
rows).

\subsection*{Receptive field diversity}

Sensory neurons display a variety of receptive field shapes \citep{Ringach_2002},
and recent modeling efforts \citep{Rehn_Sommer_2007,zylberberg_sparse_2011}
have attempted to understand the properties that give rise to the
specific receptive fields seen in experiments. We show here that the
shape diversity of a model can be predicted by our projection pursuit
analysis, and is primarily determined by the statistics of input representation,
while relatively robust to the specific effective nonlinearity.

\begin{figure}[!htbp]
\begin{centering}
\includegraphics[width=12cm]{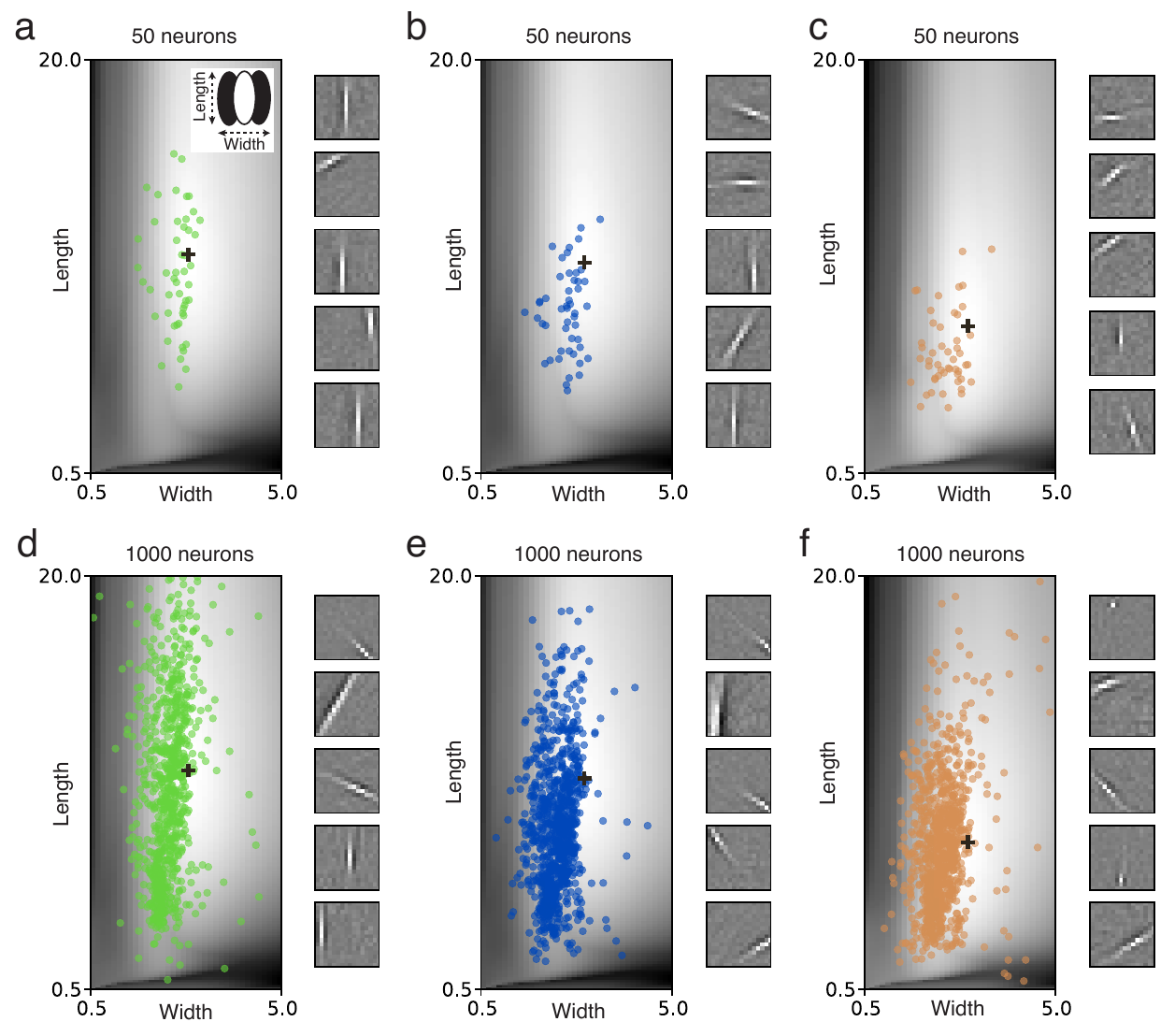}
\par\end{centering}

\caption[Optimal receptive field shapes in model networks induce diversity.]{\textbf{Optimal receptive field shapes in model networks induce diversity.}
(\textbf{a-f}) Gray level indicates the optimization value for different
lengths and widths (see inset in \textbf{a}) of oriented receptive
fields for natural images, for the quadratic rectifier (left, see
Fig.  \ref{fig:winners}a), linear rectifier (middle) and
$L_{0}$ sparse coding (right). Optima marked with a black cross.
(\textbf{a-c}) Colored circles indicate the receptive fields of different
shapes developed in a network of 50 neurons with lateral inhibitory
connections. Insets on the right show example receptive fields developed
during simulation. (\textbf{d-f}) Same for a network of 1000 neurons. }
\label{fig:gabormap}
\end{figure}

We studied a model with multiple neurons in the second layer, which
compete with each other for the representation of specific features
of the input. Each neuron had a piece-wise linear f-I curve and a
quadratic rectifier plasticity function (see Methods) and projected
inhibitory connections $v$ onto all others. These inhibitory connections
are learned by anti-Hebbian plasticity and enforce decorrelation of
neurons, so that receptive fields represent different positions, orientations
and shapes \citep{foldiak_forming_1990,vogels_inhibitory_2011,King_Zylberberg_DeWeese_2013}.
For 50 neurons, the resulting receptive fields became diversified
(Fig. \ref{fig:gabormap}a-c, colored dots). In an overcomplete
network of 1000 neurons, the diversity further increased (Fig. \ref{fig:gabormap}d-f, colored dots). 

For the analysis of the simulation results, we refined our inspection
of optimal oriented receptive fields for natural images by numerical
evaluation of the optimality criterion $\langle F(\mathbf{w}^{T}\mathbf{x})\rangle$
for receptive fields \textbf{$\mathbf{w}=\mathbf{w}_{Gabor}$}, described
as Gabor functions of variable length, width and spatial frequency.
For all tested nonlinearities, the optimization function for single-neuron
receptive fields varies smoothly with these parameters (Fig
\ref{fig:gabormap}, grey-shaded background). The single-neuron optimality
landscape was then used to analyze the multi-neuron simulation results.
We found that receptive fields are located in the area where the single-neuron
optimality criterion is near its maximum, but spread out so as to
represent different features of the input (Fig. \ref{fig:gabormap}).
Thus the map of optimization values, calculated from the theory of
effective nonlinearity, enables us to qualitatively predict the shape
diversity of receptive fields. 

\begin{figure}[!htbp]
\begin{centering}
\includegraphics{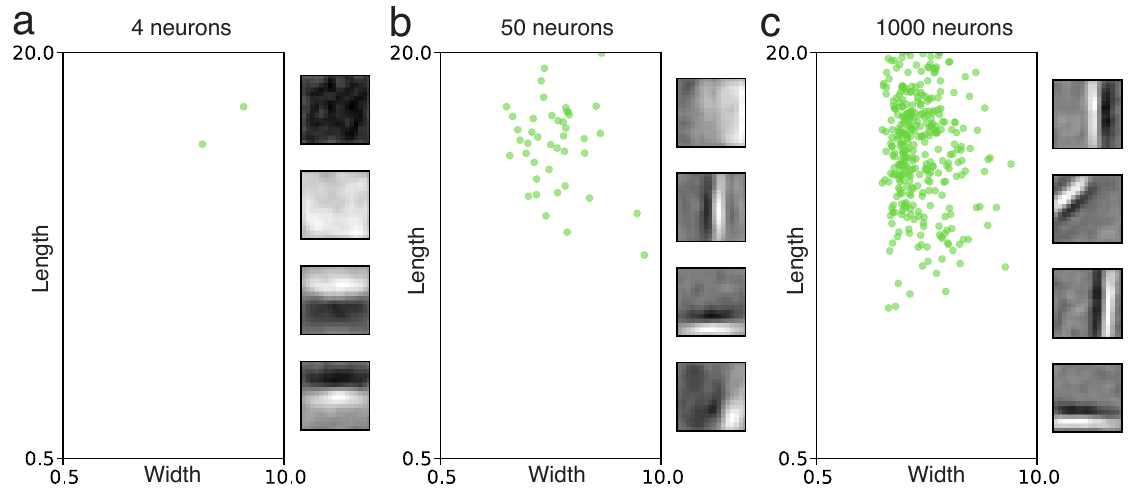}
\par\end{centering}

\caption[Receptive fields for non-whitened natural images.]{\textbf{Receptive fields for non-whitened natural images.}Images were preprocessed
as in the original sparse coding study \citep{olshausen_sparse_1997}.
We simulated linear rectifier neurons ($\theta=0.5)$ with a quadratic
plasticity nonlinearity ($b=0.5$). (\textbf{a}) Multiple-neuron simulations,
with 4 neurons. The principal components dominate the optimization
and receptive fields are not local, since they extend over most of
the image patch.\textbf{ }With 50 (\textbf{b}) and 1000 (\textbf{c})
neurons, lateral inhibition promotes diversity, and more localized
receptive field are formed. (\textbf{insets}) Sample receptive fields
developed for each simulation.}
 \label{fig:nonwhitened}
\end{figure}

Although qualitatively similar, there are differences in the receptive
fields developed for each model, such as smaller lengths for the $L_{0}$
sparse coding model (Fig. \ref{fig:gabormap}c). While potentially
significant, these differences across models may be overwhelmed by
differences due to other model properties, including different network
sizes or input representations. This is illustrated by observing that
receptive field diversity for a given model differ substantially across
network sizes (Fig. \ref{fig:gabormap}), and the difference
is even greater from simulations with an input that is not completely
white (Fig. \ref{fig:nonwhitened}c). Thus our results suggests
that efforts to model receptive field shapes observed experimentally
\citep{Ringach_2002,Rehn_Sommer_2007,zylberberg_sparse_2011} should
focus on network size and input representation, which potentially
have a stronger effect than the nonlinear properties of the specific
model under consideration.

\begin{figure}[!htbp]
\begin{centering}
\includegraphics{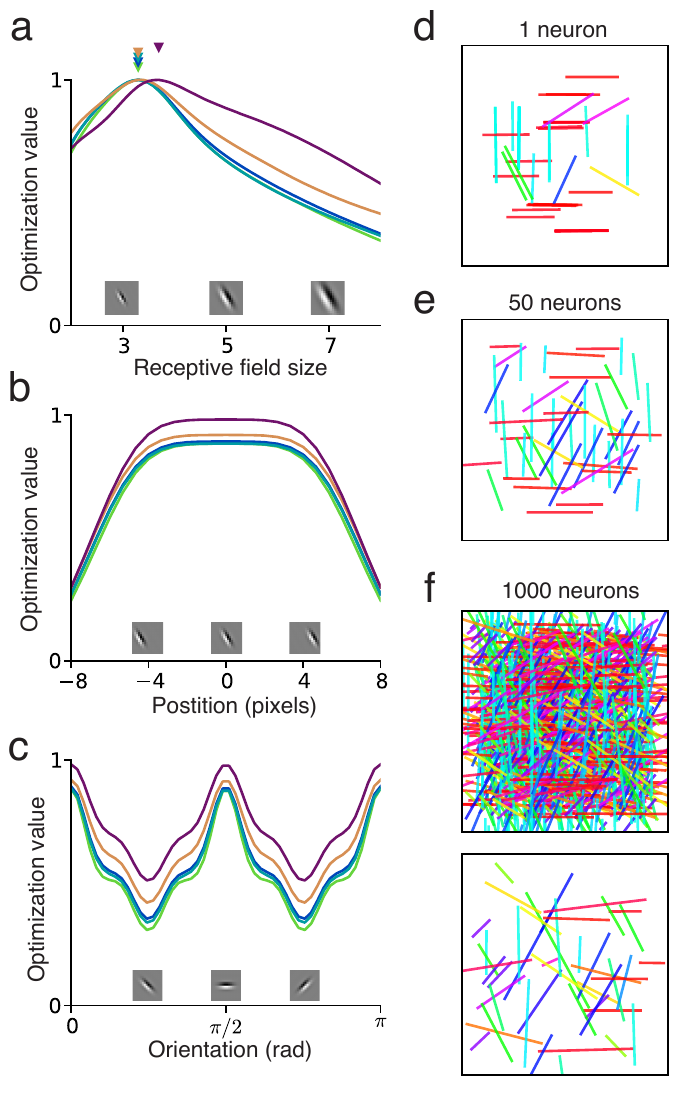}
\par\end{centering}

\caption[Diversity of receptive field size, position and orientation.]{\textbf{Diversity of receptive field size, position and orientation.} (\textbf{a})
The optimization value of localized oriented receptive fields, within
a 16x16 pixel patch of sensors, as a function of size (see Methods),
for five nonlinearities (colors as in Fig. \ref{fig:winners}a).
Optimal size is a receptive field of width around 3 to 4 pixels (filled
triangles). (\textbf{b}) The optimization value as a function of position
of the receptive field center, for a receptive field width of 4 pixels,
indicates invariance to position within the 16x16 patch, except near
the borders. (\textbf{c}) The optimization value as a function of
orientation shows preference toward horizontal and vertical directions,
for all five nonlinearities.\textbf{ }(\textbf{d}) Receptive field
position, orientation and length (colored bars) learned for 50 single-neuron
trials. The color code indicates different orientations. (\textbf{e})
Receptive field positions and orientations learned in a 50 neuron
network reveal diversification of positions, except at the borders.
(\textbf{f}) With 1000 neurons, positions and orientations cover the
full range of combinations (top). Selecting 50 randomly chosen receptive
fields highlights the diversification of position, orientation and
size (bottom). Receptive fields were learned through the quadratic
rectifier nonlinearity ($\theta_{1}=1.$, $\theta_{2}=2.$).}

\label{fig:position}
\end{figure}

We also studied the variation of receptive field position and orientation.
For all five nonlinearities considered, the optimization value is
equal for different positions of the receptive field centers, confirming
the translation invariance in the image statistics, as long as the
receptive field is not too close to the border of the anatomically
allowed fan-in of synaptic connections (Fig. \ref{fig:position}b).
Also, all nonlinearities reveal the same bias towards the horizontal
and vertical orientations (Fig. \ref{fig:position}c). These
optimality predictions are confirmed in single neuron simulations,
which lead mostly to either horizontal or vertical orientations, at
random positions (Fig. \ref{fig:position}d). When the network
is expanded to 50 neurons, recurrent inhibition forces receptive fields
to cover different positions, though excluding border positions, and
some neurons have non-cardinal orientations (Fig. \ref{fig:position}e).
With 1000 neurons, receptive fields diversify to many possible combinations
of position, orientation and length (Fig. \ref{fig:position}f). 

\subsection*{Beyond V1 simple cells}

Nonlinear Hebbian learning is not limited to explaining simple cells
in V1. We investigated if the same learning principles apply to receptive
field development in other visual or auditory areas or under different
rearing conditions. 

\begin{figure}[!htbp]
\begin{centering}
\includegraphics[height=0.6\textheight]{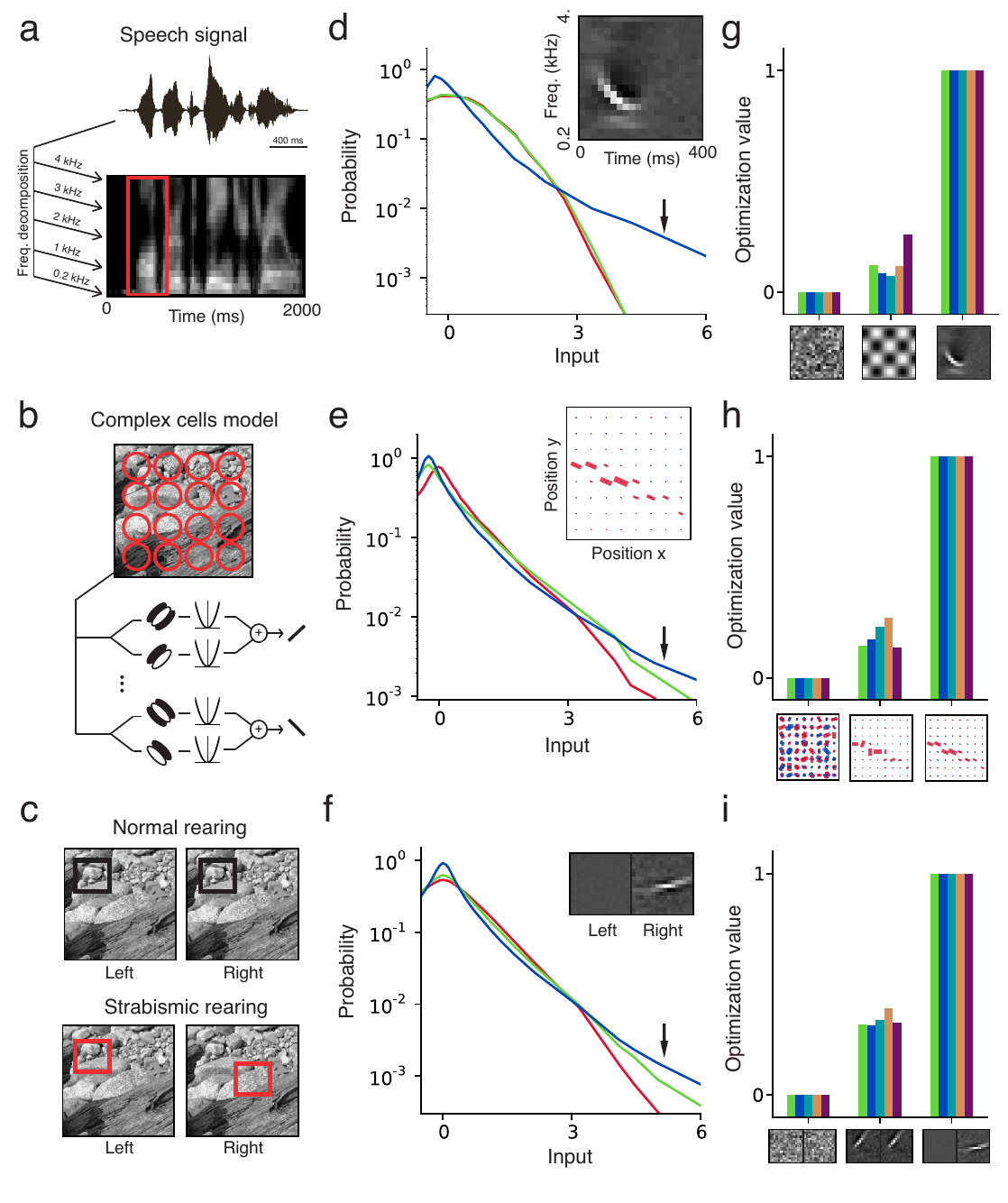}
\par\end{centering}

\caption[Nonlinear Hebbian learning across sensory modalities.]{\textbf{Nonlinear Hebbian learning across sensory modalities.} (\textbf{a})
The auditory input is modeled as segments over time and frequency
(red) of the spectrotemporal representation of speech signals. (\textbf{b})
The V2 input is assembled from the output of modeled V1 complex cells
at different positions and orientations. Receptive fields are represented
by bars with size proportional to the connection strength to the complex
cell with the respective position and orientation. (\textbf{c}) Strabismic
rearing is modeled as binocular stimuli with non-overlapping left
and right eye input patches (red). (\textbf{d-f}) Statistical distribution
(log scale) of the input projected onto three different features for
speech (\textbf{d}), V2 (\textbf{e}) and strabismus (\textbf{f}).
In all three cases, the learned receptive field (blue, inset) is characterized
by a longer tailed distribution (arrows) than the random (red) and
comparative (green) features. (\textbf{g-i})\textbf{ }Relative optimization
value for five nonlinearities (same as in Fig. \ref{fig:winners}),
for the three selected patterns (\textbf{insets}). The receptive fields
learned with the quadratic rectifier nonlinearity ($\theta_{1}=1.$,
$\theta_{2}=2.$) are the maxima among the three patterns, for all
five nonlinearities, for all three datasets.}
\label{fig:othermodalities}
\end{figure}

For auditory neurons \citep{Smith_Lewicki_2006}, we used segments
of speech as input (Fig. \ref{fig:othermodalities}a) and
observed the development of spectrotemporal receptive fields localized
in both frequency and time \citep{Miller_Escabi_Read_Schreiner_2002}
(Fig. \ref{fig:othermodalities}d). The statistical distribution
of input patterns aligned with the learned receptive fields had longer
tails than for random or non-local receptive fields, indicating temporal
sparsity of responses (Fig. \ref{fig:othermodalities}d).
Similar to our simple cell results, the learned receptive fields show
higher optimization value for all five effective nonlinearities (Fig
\ref{fig:othermodalities}g).

For a study of receptive field development in the secondary visual
cortex (V2)  \citep{Lee_Ekanadham_Ng_2007}, we used natural images
and the standard energy model \citep{Hyvarinen_Hurri_Hoyer_2009} of
V1 complex cells to generate input to V2 (Fig. \ref{fig:othermodalities}b).
The learned receptive field was selective to a single orientation
over neighboring positions, indicating a higher level of translation
invariance. When inputs were processed with this receptive field,
we found longer tails in the feature distribution than with random
features or receptive fields without orientation coherence (Fig
\ref{fig:othermodalities}e), and the learned receptive field had
a higher optimization value for all choices of nonlinearity (Fig
\ref{fig:othermodalities}h).

Another important constraint for developmental models are characteristic
deviations, such as strabismus, caused by abnormal sensory rearing.
Under normal binocular rearing conditions, the fan-in of synaptic
input from the left and right eyes overlap in visual space (Fig
\ref{fig:othermodalities}c). In this case, binocular receptive fields
with similar features for left and right eyes develop. In the strabismic
condition, the left and right eyes are not aligned, modeled as binocular
rearing with non-overlapping input from each eye (Fig. \ref{fig:othermodalities}c).
In this scenario, a monocular simple cell-like receptive field developed
(Fig. \ref{fig:othermodalities}f), as observed in experiments
and earlier models \citep{Cooper_Intrator_Blais_Shouval_2004}. The
statistical distributions confirm that for disparate inputs the monocular
receptive field is more kurtotic than a binocular one, explaining
its formation in diverse models \citep{Hunt_Dayan_Goodhill_2013} (Fig
\ref{fig:othermodalities}f,i).

Our results demonstrate the generality of the theory across multiple
cortical areas. Selecting a relevant feature space for an extensive
analysis, as we have done with simple cells and natural images, may
not be possible in general. Nonetheless, nonlinear Hebbian learning
helps to explain why some features (and not others) are learnable
in network models \citep{Saxe_Bhand_Mudur_Suresh_Ng_2011}. 

\section*{Discussion}

Historically, a variety of models have been proposed to explain the
development and distribution of receptive fields. We have shown that
nonlinear Hebbian learning is a parsimonious principle which is implicitly
or explicitly present in many developmental models \citep{olshausen_emergence_1996,bell_independent_1997,law_formation_1994,Rehn_Sommer_2007,clopath_connectivity_2010,Savin_Joshi_Triesch_2010,zylberberg_sparse_2011,pfister_triplets_2006,hyvarinen_independent_1998,foldiak_forming_1990,Hunt_Dayan_Goodhill_2013}.
The fact that receptive field development is robust to the specific
nonlinearity highlights a functional relation between different models.
It also unifies feature learning across sensory modalities: receptive
fields form around features with a long-tailed distribution.

\subsection*{Relation to previous studies}

Earlier studies have already placed developmental models side by side,
comparing their normative assumptions, algorithmic implementation
or receptive fields developed. Though consistent with their findings,
our results lead to revised interpretations and predictions.

The similarities between sparse coding and ICA are clear from their
normative correspondence \citep{olshausen_sparse_1997}. Nevertheless,
the additional constraint in ICA, of having at most as many features
as inputs, makes it an easier problem to solve, allowing for a range
of suitable algorithms \citep{hyvarinen_independent_2000}. These differ
from algorithms derived for sparse coding, in which the inference
step is difficult due to overcompleteness. We have shown that regardless
of the specific normative assumptions, it is the common implementation
of nonlinear Hebbian learning that explains similarities in their
learning properties. 

In contrast to the idea that in sparse coding algorithms overcompleteness
is required for development of localized oriented edges \citep{olshausen_sparse_1997},
we have demonstrated that a sparse coding model with a single neuron
is mathematically equivalent to nonlinear Hebbian learning and learns
localized filters in a setting that is clearly \textquotedbl{}undercomplete\textquotedbl{}.
Thus differences observed in receptive field shapes between sparse
coding and ICA models \citep{Ringach_2002} are likely due to differences
in network size and input preprocessing. For instance, the original
sparse coding model \citep{olshausen_sparse_1997} applied a preprocessing
filter that did not completely whiten the input, leading to larger
receptive fields (Fig. \ref{fig:nonwhitened}).

Studies that derive spiking models from normative theories often interpret
the development of oriented receptive fields as a consequence of its
normative assumptions \citep{Savin_Joshi_Triesch_2010,zylberberg_sparse_2011}.
In a recent example, a spiking network has been related to the sparse
coding model \citep{zylberberg_sparse_2011}, using neural properties
defined ad hoc. Our results suggest that many other choices of neural
activations would have given qualitatively similar receptive fields,
independent of the sparse coding assumption. While in sparse coding
the effective nonlinearity derives from a linear plasticity rule combined
with a nonlinear f-I curve, our results indicate that a nonlinear
plasticity rule combined with a linear neuron model would give the
same outcome.

In order to distinguish between different normative assumptions, or
particular neural implementations, the observation of \textquotedbl{}oriented
filters\textquotedbl{} is not sufficient and additional constraints
are needed. Similarly receptive shape diversity, another important
experimental constraint, should also be considered with care, since
it cannot easily distinguish between models either. Studies that confront
the receptive field diversity of a model to experimental data \citep{Rehn_Sommer_2007,zylberberg_sparse_2011,Ringach_2002}
should also take into account input preprocessing choices and how
the shape changes with an increasing network size, since we have observed
that these aspects may have a larger effect on receptive field shape
than the particulars of the learning model.

Empirical studies of alternative datasets, including abnormal visual
rearing \citep{Hunt_Dayan_Goodhill_2013}, tactile and auditory stimuli
\citep{Saxe_Bhand_Mudur_Suresh_Ng_2011}, have also observed that different
unsupervised learning algorithms lead to comparable receptive fields
shapes. Our results offer a plausible theoretical explanation for
these findings. 

Past investigations on nonlinear Hebbian learning \citep{Fyfe_Baddeley_1995,hyvarinen_independent_1998}
demonstrated that many nonlinearities were capable of solving the
cocktail party problem. Since it is a specific toy model, that asks
for the unmixing of linearly mixed independent features, it is not
clear a priori whether the same conclusions would hold in other settings.
We have shown that the results of \citet{Fyfe_Baddeley_1995} and \citet{hyvarinen_independent_1998}
generalize in two directions. First, the effective nonlinear Hebbian
learning mechanism is also behind other models beyond ICA, such as
sparse coding models and plastic spiking networks. Second, the robustness
to the choice of nonlinearity is not limited to a toy example, but
also holds in multiple real world data. Together, these insights explain
and predict the outcome of many developmental models, in diverse applications.

\subsection*{Robustness to normative assumptions}

Many theoretical studies start from normative assumptions \citep{bell_independent_1997,Rehn_Sommer_2007,Savin_Joshi_Triesch_2010,olshausen_sparse_1997},
such as a statistical model of the sensory input or a functional objective,
and derive neural and synaptic dynamics from them. Our claim of universality
of feature learning indicates that details of normative assumptions
may be of lower importance.

For instance, in sparse coding one assumes features with a specific
statistical prior \citep{Rehn_Sommer_2007,olshausen_sparse_1997}.
After learning, this prior is expected to match the posterior distribution
of the neuron's firing activity \citep{Rehn_Sommer_2007,olshausen_sparse_1997}.
Nevertheless, we have shown that receptive field learning is largely
unaffected by the choice of prior. Thus, one cannot claim that the
features were learned because they match the assumed prior distribution,
and indeed in general they do not. For a coherent statistical interpretation,
one could search for a prior that would match the feature statistics.
However, since the outcome of learning is largely unaffected by the
choice of prior, such a statistical approach would have limited predictive
power. Generally, kurtotic prior assumptions enable feature learning,
but the specific priors are not as decisive as one might expect. Because
normative approaches have assumptions, such as independence of hidden
features, that are not generally satisfied by the data they are applied
to, the actual algorithm that is used for optimization becomes more
critical than the formal statistical model.

The concept of sparseness of neural activity is used with two distinct
meanings. The first one is a single-neuron concept and specifically
refers to the long-tailed distribution statistics of neural activity,
indicating a \textquotedbl{}kurtotic\textquotedbl{} distribution.
The second notion of sparseness is an ensemble concept and refers
to the very low firing rate of neurons, observed in cortical activity
\citep{barth_experimental_2012}, which may arise from lateral competition
in overcomplete representations. Overcompleteness of ensembles makes
sparse coding different from ICA \citep{olshausen_sparse_1997}. We
have shown here that competition between multiple neurons is fundamental
for receptive field diversity, whereas it is not required for simple
cell formation per se. Kurtotic features can be learned even by a
single neuron with nonlinear Hebbian learning, and with no restrictions
on the sparseness of its firing activity.

\subsection*{Interaction of selectivity with preprocessing and homeostasis}

The concept of nonlinear Hebbian learning also clarifies the interaction
of feature selectivity with preprocessing mechanisms. We have assumed
whitened data throughout the study, except Fig. \ref{fig:nonwhitened}.
Since after whitening second-order correlations are uninformative,
neurons can develop sensitivity to higher order features. While whitened
data is formally not required for our analysis, second-order correlations
may dominate the optimization for non-white input, so that principal
components will be learned (Fig. \ref{fig:nonwhitened}a).
Only when multiple neurons are added and receptive fields diversify,
are localized simple cells formed with an input that is not completely
white \citep{olshausen_sparse_1997} (Fig. \ref{fig:nonwhitened}c). 

In studies of spiking networks, the input is restricted to positive
rates, possibly through an on/off representation, as observed in the
LGN \citep{Miller_1994}. While the center-surround properties of LGN
contributes to a partial decorrelation of neuronal activity \citep{Dan_Atick_Reid_1996},
in such alternative representations, trivial receptive fields may
develop, such as a single non-zero synapse, and additional mechanisms,
such as hard bounds on each synaptic strength, $a\le w_{j}\le b$,
may be necessary to restrict the optimization space to desirable features
\citep{clopath_connectivity_2010}. 

Instead of constraining the synaptic weights, one may implement a
synaptic decay as in Oja's plasticity rule \citep{Oja_1982}, $\Delta w\propto x\cdot y-w\cdot y^{2}$
(see also \citep{Chen_Lonjers_Lee_Chistiakova_Volgushev_Bazhenov_2013}).
Because of its multiplicative effect, the decay term does not alter
the receptive field, but only scales its strength. Thus, it is equivalent
to rescaling the input in the f-I curve, so as to shift it to the
appropriate range (Fig. \ref{fig:selectivity}). Similar scaling
effects arise from f-I changes due to intrinsic plasticity \citep{Savin_Joshi_Triesch_2010,turrigiano_too_2011,Elliott_2014}.
The precise relation between nonlinear Hebbian learning, spiking representations
and homeostasis in the cortex is an important topic for further studies.

\subsection*{Universality supports biological instantiation}

The principle of nonlinear Hebbian learning has a direct correspondence
to biological neurons and is compatible with a large variety of plasticity
mechanisms. It is not uncommon for biological systems to have diverse
implementations with comparable functional properties \citep{Prinz_Bucher_Marder_2004}.
Different species, or brain areas, could have different neural and
plasticity characteristics, and still have similar feature learning
properties \citep{Sharma_Angelucci_Sur_2000,Kaschube_Schnabel_Lowel_Coppola_White_Wolf_2010}.
The generality of the results discussed in this paper reveals learning
simple cell-like receptive fields from natural images to be much easier
than previously thought. It implies that a biological interpretation
of models is possible even if some aspects of a model appear simplified
or even wrong in some biological aspects. Universality also implies
that the study of receptive field development is not sufficient to
distinguish between different models.

The relation of nonlinear Hebbian learning to projection pursuit endorses
the interpretation of cortical plasticity as an optimization process.
Under the rate coding assumptions considered here, the crucial property
is an effective synaptic change linear in the pre-synaptic rate, and
nonlinear in the post-synaptic input. Pairing experiments with random
firing and independently varying pre- and post-synaptic rates would
be valuable to investigate these properties \citep{Sjostrom_Turrigiano_Nelson_2001,Sjostrom_Rancz_Roth_Hausser_2008,Graupner_Brunel_2012}.
Altogether, the robustness to details in both input modality and neural
implementation suggests nonlinear Hebbian learning as a fundamental
principle underlying the development of sensory representations.

\section*{Methods}

\textbf{Spiking model.} A generalized leaky integrate-and-fire neuron
\citep{pozzorini_temporal_2013} was used as spiking model, which includes
power-law spike-triggered adaptation and stochastic firing, with parameters
\citep{pozzorini_temporal_2013} fitted to pyramidal neurons. The f-I
curve $g(I)$ was estimated by injecting step currents and calculating
the trial average of the spike count over the first $500$ ms. The
minimal triplet-STDP model\citep{pfister_triplets_2006} was implemented,
in which synaptic changes follow 

\begin{equation}
\frac{d}{dt}w(t)=A^{+}y(t)\bar{y}^{+}(t)\bar{x}^{+}(t)-A^{-}x(t)\bar{y}^{-}(t)
\end{equation}
where $y(t)$ and $x(t)$ are the post- and pre-synaptic spike trains,
respectively: $y(t)=\sum_{f}\delta(t-t^{f})$, where $t^{f}$ are
the firing times and $\delta$ denotes the Dirac $\delta$-function;
$x(t)$ is a vector with components $x_{i}(t)=\sum_{f}\delta(t-t_{i}^{f})$,
where $t_{i}^{f}$ are the firing times of pre-synaptic neuron $i$;
$w$ is a vector comprising the synaptic weights $w_{i}$ connecting
a pre-synaptic neuron $i$ to a post-synaptic cell. $A^{+}=6.5\cdot10^{-3}$
and $A^{-}=5.3\cdot10^{-3}$ are constants, and $\bar{y}^{+}$, $\bar{x}^{+}$
and $\bar{y}^{-}$ are moving averages, implemented by integration
(e.g. $\tau\frac{\partial\bar{{y}}}{\partial t}=-\bar{y}+y$), with
time scales $114.0$ ms, $16.8$ ms and $33.7$ ms, respectively \citep{pfister_triplets_2006}.
For estimating the nonlinearity $h(y)$ of the plasticity, pre- and
post-synaptic spike trains were generated as Poisson processes, with
the pre-synaptic rate set to $20$ Hz.

A linear rectifier $g(x)=a(x-b)_{+}$ was fitted to the f-I curve
of the spiking neuron model by squared error optimization. Similarly,
a quadratic function $h(x)=a(x^{2}-bx)$ was fitted to the nonlinearity
of the triplet STDP model. The combination of these two fitted functions
was plotted as fit for the effective nonlinearity $f(x)=h(g(x))$.

\textbf{Sparse coding analysis.} A sparse coding model, with $K$
neurons $y_{1},\dots,y_{K}$, has a nonlinear Hebbian learning formulation.
The sparse coding model minimizes a least square reconstruction error
between the vector of inputs $\mathbf{x}$ and the reconstruction
vector $\mathbf{W}\mathbf{y},$ where $\mathbf{W}=[\mathbf{w}_{1}\dots\mathbf{w}_{K}]$,
and $\mathbf{y}=(y_{1},\dots,y_{K})$ is the vector of neuronal activities,
with $y_{j}\ge0$ for $1\le j\le K$. The total error $E$ combines
a sparsity constraint $S$ with weight $\lambda$ and the reconstruction
error, $E=\frac{1}{2}||\mathbf{x}-\mathbf{W}\mathbf{y}||^{2}+\lambda\sum S(y_{k})$.
$E$ has to be minimal, averaged across all input samples, under the
constraint $y_{j}\ge0$ for all $j$. 

The minimization problem is solved by a two-step procedure. In the
first step, for each input sample, one minimizes $E$ with respect
to all hidden units $y_{j}$
\begin{equation}
\begin{aligned}\frac{d}{dy_{j}}E=0 & \iff\mathbf{w}_{j}(\mathbf{x}-\mathbf{W}\mathbf{y})-\lambda S'(y_{j})=0\\
 & \iff\mathbf{w}_{j}\mathbf{x}-\sum_{k\neq j}(\mathbf{w}_{j}^{T}\mathbf{w}_{k})y_{k}-||\mathbf{w}_{j}||^{2}y_{j}-\lambda S'(y_{j})=0\\
 & \iff y_{j}+\lambda S'(y_{j})=\mathbf{w}_{j}^{T}\mathbf{x}-\sum_{k\neq j}(\mathbf{w}_{j}^{T}\mathbf{w}_{k})y_{k}\\
 & \iff y_{j}=g(\mathbf{w}_{j}^{T}\mathbf{x}-\sum_{k\neq j}v_{jk}y_{k})
\end{aligned}
\end{equation}
where we constrained the vector $\mathbf{w}_{j}$ of synapses projecting
onto unit $y_{j}$ by $||\mathbf{w}_{j}||^{2}=1$, defined the activation
function $g(.)=T^{-1}(.)$, the inverse of $T(y)=(y+\lambda S'(y))$,
and defined recurrent synaptic weights $v_{jk}=\mathbf{w}_{j}^{T}\mathbf{w}_{k}$.
For each input sample $\mathbf{x}$, this equation shall be iterated
until convergence. The equation can be interpreted as a recurrent
neural network, where each neuron has an activation function $g$,
and the input is given by the sum of the feedforward drive $\mathbf{w}_{j}^{T}\mathbf{x}$
and a recurrent inhibition term $-\sum_{k\neq j}v_{jk}y_{k}$. To
avoid instability, we implement a smooth membrane potential $u_{j}$,
which has the same convergence point \citep{rozell_sparse_2008}
\begin{equation}
\begin{aligned} & \tau_{u}\frac{d}{dt}u_{j}(t)=-u_{j}(t)+(\mathbf{w}_{j}^{T}\mathbf{x}-\sum_{k\neq j}v_{jk}y_{k}(t))\\
 & y_{j}(t)=g(u_{j}(t))
\end{aligned}
\end{equation}
initialized with $u_{j}(t)=0$. 

The second step is a standard gradient descent implementation of the
least square regression optimization, leading to an learning rule
\[
\Delta w_{j}\propto\frac{d}{dw_{j}}E=(\mathbf{x}-\mathbf{W}^{T}\mathbf{y})\ y_{j}=\mathbf{x}\ y_{j}-\mathbf{w}_{j}\ y_{j}^{2}-\sum_{k\neq j}\mathbf{w}_{k}y_{k}y_{j}
\]

The decay term $\mathbf{w}_{j}\ y_{j}^{2}$ has no effect, since the
norm is constrained to $||\mathbf{w}_{j}||=1$ at each step. For a
single unit $y,$ the model simplifies to a nonlinear Hebbian formulation,
$\Delta\mathbf{w}\propto\mathbf{x}\ g(\mathbf{w}_{j}^{T}\mathbf{x})$.
For multiple units, it can be interpreted as projection pursuit on
an effective input, not yet represented by other neurons, $\tilde{\mathbf{x}_{j}}=\mathbf{x}-\sum_{k\neq j}\mathbf{w}_{k}y_{k}$,
which simplifies to $\Delta\mathbf{w}_{j}\propto\tilde{\mathbf{x}}_{j}\cdot g(\mathbf{w}_{j}^{T}\tilde{\mathbf{x}_{j}})$
. 

There are two non-local terms that need to be implemented by local
mechanisms so as to be biologically plausible. First, the recurrent
weights depend on the overlap between receptive fields, $\mathbf{w}_{j}^{T}\mathbf{w}_{k}$,
which is non-local. The sparse coding model assumes independent hidden
neurons, which implies that after learning neurons should be pair-wise
uncorrelated, $cov(y_{j},y_{k})=0$. As an aside we note that the
choice $v_{jk}=\mathbf{w}_{j}^{T}\mathbf{w}_{k}$ does not automatically
guarantee decorrelation. Decorrelation may be enforced through plastic
lateral connections, following an anti-Hebbian rule \citep{foldiak_forming_1990,zylberberg_sparse_2011},
$\Delta v_{jk}\propto(y_{j}-\langle y_{j}\rangle)\cdot y_{k}$, where
$\langle y_{j}\rangle$ is a moving average (we use $\tau=1000$ input
samples). Thus by substituting fixed recurrent connections by anti-Hebbian
plasticity, convergence $\Delta v_{jk}=0$ implies $cov(y_{j},y_{k})=0$.
While this implementation does not guarantee $v_{jk}=\mathbf{w}_{j}^{T}\mathbf{w}_{k}$
after convergence, neither does $v_{jk}=\mathbf{w}_{j}^{T}\mathbf{w}_{k}$
guarantee decorrelation $cov(y_{j},y_{k})=0$, it does lead to optimal
decorrelation, which is the basis of the normative assumption. Additionally
we constrain $v_{jk}\geq0$ to satisfy Dale's law. Although some weights
would converge to negative values otherwise, most neuron pairs have
correlated receptive fields, and thus positive recurrent weights. 

Second, we ignore the non-local term $\sum_{k\neq j}\mathbf{w}_{k}y_{k}y_{j}$
in the update rule. Although this approximation is not theoretically
justified, we observed in simulations that receptive fields do not
qualitatively differ when this term is removed.

The resulting Hebbian formulation can be summarized as
\begin{equation}
\begin{aligned} & y_{j}=g(\mathbf{w}_{j}^{T}\mathbf{x}-\sum_{k\neq j}v_{jk}y_{k})\\
 & \Delta\mathbf{w}_{j}\propto\mathbf{x\ }y_{j}\\
 & \Delta v_{jk}\propto(y_{j}-\langle y_{j}\rangle)\cdot y_{k}
\end{aligned}
\label{eq:sc_hebb}
\end{equation}

This derivation unifies previous results on the biological implementation
of sparse coding: the relation of the sparseness constraint to a specific
activation function \citep{rozell_sparse_2008}, the derivation of
a Hebbian learning rule from quadratic error minimization \citep{Oja_1982},
and the possibility of approximating lateral interaction terms by
learned lateral inhibition \citep{foldiak_forming_1990,zylberberg_sparse_2011}.

\textbf{Nonlinearities and optimization value. }The optimization value
for a given effective nonlinearity $f$, synaptic weights $w$, and
input samples $x$, is given by $R=\langle F(\mathbf{w}^{T}\mathbf{x})\rangle$,
where $F=\int f$ and angular brackets indicate the ensemble average
over $x$. Relative optimization values in Figs. \ref{fig:winners}b
and \ref{fig:position} were normalized to $[0,1]$, relative
to the minimum and maximum values among the considered choice of features
$w$, $R^{*}=(R-R_{min})/(R_{max}-R_{min})$.\textbf{ }The selectivity
index of a nonlinearity $f$ is defined as $SI=(\langle F(l)\rangle-\langle F(g)\rangle)/\sigma_{F}$,
where $l$ and $g$ are Laplacian and Gaussian variables respectively,
normalized to unit variance. $\sigma_{F}=\sqrt{\sigma_{F(l)}\sigma_{F(g)}}$
is a normalization factor, with \textrm{$\sigma_{F(.)}=\sqrt{\langle F(.)^{2}\rangle}$}.
The selectivity of an effective nonlinearity $f$ is not altered by
multiplicative scaling, $\tilde{f}(u)=\alpha f(u)$, neither by additive
constants when the input distribution is symmetric, $\tilde{f}(u)=\alpha f(u)+\beta$.
The effective nonlinearities in Fig. \ref{fig:winners} included
the linear rectifier $f(u)=\begin{cases}
0, & if\ u<\theta\\
u-\theta, & if\ u\ge\theta
\end{cases}$, the quadratic rectifier $f(u)=\begin{cases}
0, & if\ u<\theta\\
(u-\theta)(u-\theta-b), & if\ u\ge\theta
\end{cases}$, the $L_{0}$ sparse coding nonlinearity $f(u)=\begin{cases}
0, & if\ u<\lambda\\
u, & if\ u\ge\lambda
\end{cases}$, the Cauchy sparse coding nonlinearity $f=T^{-1}$, where $T(y)=\begin{cases}
0, & if\ y<0\\
y+2\lambda y/(1+y^{2}), & if\ y\ge0
\end{cases}$, the negative sigmoid $f(u)=1-2/(1+e^{-2u})$, a polynomial function
$f(u)=u^{3}$, trigonometric functions $sin(u)$ and \textbf{$cos(u)$},
a symmetric piece-wise linear function $f(u)=\begin{cases}
0, & if\ |u|<\theta\\
|u|-\theta, & if\ |u|\ge\theta
\end{cases}$, as well as, for comparison, a linear function $f(u)=u$.

\textbf{Receptive field learning. }Natural image patches (16 by 16
pixel windows) were sampled from a standard dataset \citep{olshausen_emergence_1996}
($10^{6}$ patches). Patches were randomly rotated by $\pm90{}^{\circ}$
degrees to avoid biases in orientation. The dataset was whitened by
mean subtraction and a standard linear transformation $\mathbf{x}^{*}=\mathbf{M}\mathbf{x}$,
where $\mathbf{M}=\mathbf{R}\mathbf{D}^{-1/2}\mathbf{R}^{T}$ and
$\langle\mathbf{x}\mathbf{x}^{T}\rangle=\mathbf{R}\mathbf{D}\mathbf{R}^{T}$
is the eigenvalue decomposition of the input correlation matrix. In
Fig. \ref{fig:nonwhitened}, we used images preprocessed as
in \citet{olshausen_emergence_1996}, filtered in the spatial frequency
domain by $M(f)=f\ e^{-(f/f_{0})^{4}}$. The exponential factor is
a low-pass filter that attenuates high-frequency spatial noise, with
$f_{0}=200$ cycles per image. The linear factor $f$ was designed
to whiten the images by canceling the approximately $1/f$ power law
spatial correlation observed in natural images \citep{ruderman_statistics_1994}.
But since the exponent of the power law for this particular dataset
has an exponent closer to $1.2$, the preprocessed image exhibit higher
variance at lower spatial frequencies.

Synaptic weights were initialized randomly (normal distribution with
zero mean) and, for an effective nonlinearity $f$, evolved through
$\mathbf{w}_{k+1}=\mathbf{w}_{k}+\eta\ \mathbf{x}\ f(\mathbf{w}_{k}^{T}\mathbf{x}_{k})$,
for each input sample $x_{k}$, with a small learning rate $\eta$.
We enforced normalized weights at each time step, $||\mathbf{w}||_{2}=1$,
through multiplicative normalization, implicitly assuming rapid homeostatic
mechanisms \citep{turrigiano_too_2011,Zenke_Hennequin_Gerstner_2013}.
For multiple neurons, the neural version of the sparse coding model
described in Eq \ref{eq:sc_hebb} was implemented. In Fig
\ref{fig:gabormap} and \ref{fig:nonwhitened}, the learned
receptive fields were fitted to Gabor filters by least square optimization.
Receptive fields with less than $0.6$ variance explained were rejected
(less than 5\% of all receptive fields). 

\textbf{Receptive field selection.} In Fig. \ref{fig:winners}b,
the five selected candidate patterns are: random connectivity filter
(weights sampled independently from the normal distribution with zero
mean), high-frequency Fourier filter (with equal horizontal and vertical
spatial periods, $T_{x}=T_{y}=8$ pixels), difference of Gaussians
filter ($\sigma_{1}=3.$, $\sigma_{2}=4.$), low-frequency Fourier
filter ($T_{x}=16$, $T_{y}=32$), and centered localized Gabor filter
($\sigma_{x}=1.5$, $\sigma_{y}=2.0$, $f=0.2$, $\theta=\pi/3$,
$\phi=\pi/2$). Fourier filters were modeled as $w_{ab}=sin(2\pi a/T_{x})*cos(2\pi b/T_{y})$;
difference of Gaussians filters as the difference between two centered
2D Gaussians with same amplitude and standard deviations $\sigma_{1}$
and $\sigma_{2}$; and we considered standard Gabor filters, with
center $(x_{c},y_{c}),$ spatial frequency $f$, width $\sigma_{x}$,
length $\sigma_{y},$ phase $\phi$ and angle $\theta$. In Fig
\ref{fig:gabormap} and \ref{fig:nonwhitened} we define
the Gabor width and length in pixels as 2.5 times the standard deviation
of the respective Gaussian envelopes, $\sigma_{x}$ and $\sigma_{y}$.
In Fig. \ref{fig:position}a, a Gabor filter of size $s$
had parameters $\sigma_{x}=0.3\cdot s$, $\sigma_{y}=0.6\cdot s$,
$f=1/s$ and $\theta=\pi/3$. In Fig. \ref{fig:position}b-c,
the Gabor filter parameters were $\sigma_{x}=1.2$, $\sigma_{y}=2.4$,
$f=0.25$. All receptive fields were normalized to $||\mathbf{w}||_{2}=1$.
In Fig. \ref{fig:gabormap} and \ref{fig:nonwhitened}, the
background optimization value was calculated for Gabor filters of
different widths, lengths, frequencies, phases $\phi=0$ and $\phi=\pi/2$.
For each width and length, the maximum value among frequencies and
phases was plotted. 

\textbf{Additional datasets. }For the strabismus model, two independent
natural image patches were concatenated, representing non-overlapping
left and right eye inputs, forming a dataset with 16 by 32 patches
\citep{Cooper_Intrator_Blais_Shouval_2004}. For the binocular receptive
field in the strabismus statistical analysis (Fig. \ref{fig:othermodalities}a),
a receptive field was learned with a binocular input with same input
from left and right eyes. As V2 input, V1 complex cell responses were
obtained from natural images as in standard energy models \citep{Hyvarinen_Hurri_Hoyer_2009},
modeled as the sum of the squared responses of simple cells with alternated
phases. These simple cells were modeled as linear neurons with Gabor
receptive fields ($\sigma_{x}=1.2$, $\sigma_{y}=2.4$, $f=0.3$),
with centers placed on a 8 by 8 grid (3.1 pixels spacing), with 8
different orientations at each position (total of 512 input dimensions).
For the non-orientation selective receptive field in the V2 statistical
analysis (Fig. \ref{fig:othermodalities}d), the orientations
of the input complex cells for the learned receptive field were randomized.
As auditory input, spectrotemporal segments were sampled from utterances
spoken by a US English male speaker (CMU US BDL ARCTIC database, \citet{Kominek_Black_2004}).
For the frequency decomposition \citep{Smith_Lewicki_2006}, each audio
segment was filtered by gammatone kernels, absolute and log value
taken and downsampled to $50$ Hz. Each sample was 20 time points
long ($400$ ms segment) and 20 frequency points wide (equally spaced
between $0.2$ kHz and $4.0$ kHz). For the non-local receptive field
in the auditory statistical analysis (Fig. \ref{fig:othermodalities}g),
a Fourier filter was used ($T_{t}=T_{f}=10$). For all datasets, the
input ensemble was whitened after the preprocessing steps, by the
same linear transformation described above for natural images, and
all receptive fields were normalized to $||\mathbf{w}||_{2}=1$.
 
\section*{Acknowledgments} 

We thank C. Pozzorini and J. Brea for valuable comments, and D.S. Corneil for critical reading of the manuscript. This research was supported by the European Research Council under grant agreement no. 268689 (MultiRules). 
 
\bibliographystyle{unsrtnat}
\bibliography{nonlinear}

\end{document}